\documentclass{article}
\pdfoutput=1
\usepackage{amsfonts}
\usepackage{amsmath}
\usepackage{xspace}
\usepackage[top=1in, bottom=1.25in, left=1.1in, right=1.1in]{geometry}
\usepackage{pdfpages}

\newcommand{\bs}{\boldsymbol}
\newcommand{\bbP}{\mathbb{P}}
\newcommand{\bbQ}{\mathbb{Q}^{-1}}

\begin{document}
\includepdf[pages={1-last}]{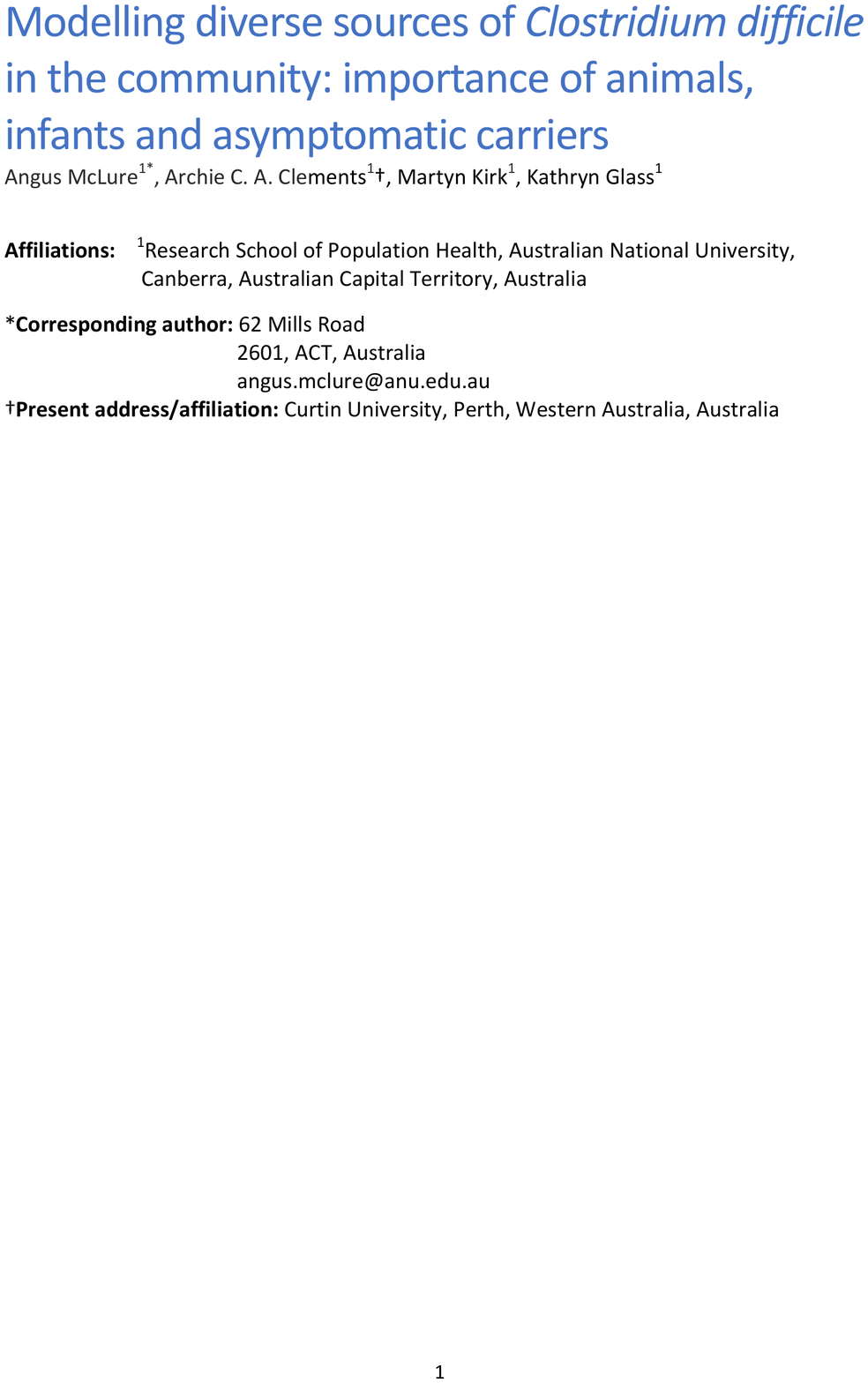}
\includepdf[pages={1-last}]{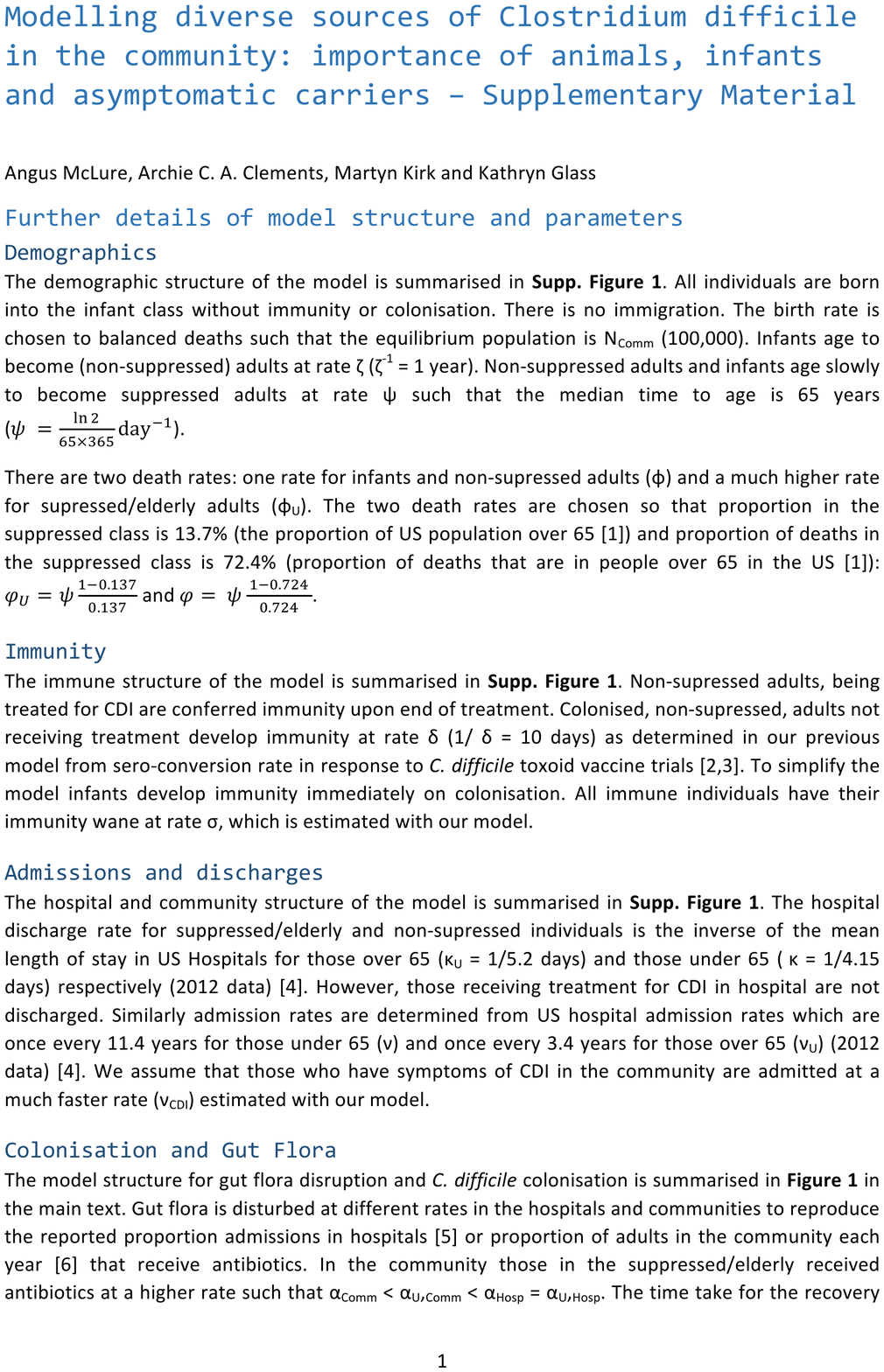}

\section*{Details of Classifying Cases as Hospital or Community-Acquired}
The simulation of CDI classification is a key component of the current study. The definitions  used to classify CDIs rely on a knowledge of the patient's history of previous CDI (to rule out recurrent cases) hospital admission and discharge prior to onset of symptoms (to distinguish hospital and community acquired cases). However, standard compartmental epidemiological models are memoryless and do not explicitly model (or record) the events occurring to individuals, only how an event (i.e. infection or recovery) affects the total number of individuals of any given compartment. An individual-based model could be used to simulate these details, but we employ a computationally much simpler approximation that approximates individual in the population as a simple Markov chain.

To make the individual Markov assumption, the only non-linear interaction term, the force of colonisation, is fixed as a constant in time (equal to the mean force of colonisation at equilibrium) so the population can be viewed as a large ensemble of independent Markov chains (individual people). The state space of the Markov chains is the union of $S$, the set of living states  corresponding to the compartments of the model and $\Delta$ the death state (which is the only absorbing state). New Markov chains are initialised (births and immigration) at the jump times of a Poisson process of rate $\mu(t)$ (the birth/immigration rate) which we will assume is independent of the individuals in the model and is homogeneous (i.e. a constant birth rate). At birth these Markov chains are randomly assigned to one of the living states in $S$ (the compartments in the model) according to the vector of probabilities $\bs{\pi}$. In our model there is no immigration and all births are in the same state, so $\bs{\pi}$ is the standard basis vector corresponding to the non-immune, non-colonised infant class. If there was immigration in our model, $\bs{\pi}$ would describe the probability that a new arrival was of any given compartment. These Markov chains then progress through the transient states in $S$, before being absorbed into the death state. For each individual Markov chain, the age $a$ is the time since it was initialised. So if $Q$ is the transition rate matrix between states in $S$, then the vector of probabilities that a given individual is in each state at age $a$ is
\begin{equation}
\left[P(X(a) = s)\right]_{s\in S} = e^{Qa} \bs\pi.
\end{equation}
In a whole population of these Markov chains starting with $\bs{x_0}$ people in each (living) state at time $0$, the expected number of people alive in each state at time $t$ is given by the vector
\begin{equation}
E[\bs{x}(t)]
= e^{Qt} \bs{x_0} + \int_0^t \mu(t-a) e^{Qa} \bs{\pi} da.
\end{equation}
If the birthrate $\mu$ is constant over time then
\begin{equation}
E[\bs{x}(t)]
= e^{Qt} \bs{x_0} + \mu Q^{-1} \left( e^{Qt} - I\right) \bs{\pi}.
\end{equation}
Since $Q$ is the transition matrix for transient states (i.e. there are no immortal states), the limit for large $t$ when the system approaches its equilibrium distribution is
\begin{equation}
\lim_{t\to \infty} E[\bs{x}(t)]
= - Q^{-1} \bs{\pi} \mu
\end{equation}
which is independent of $t$ and $\bs{x_0}$. The total population is then sum over the vector, $-\bs{1}^TQ^{-1} \bs{\pi} \mu$. Note that $-\bs{1}^TQ^{-1} \bs{\pi}$ is the total of the mean dwelling time in each state (i.e. the total average time each person spends alive) and so is equal to the mean life expectancy of individuals, $L$.

We have an expression for the number of individuals of a given type at equilibrium but we need to know the number individuals at equilibrium which have a given history of CDI and hospitalisation. So we are interested in the number of people at equilibrium which have (or have not) been in some set of states $S_1\subset S$, in the past $T_1$ units of time. For instance we may be interested in the population which haven't been in the hospital states ($S_1$) in the last 12 weeks ($T_1$). So consider the vector of probabilities that a individual Markov chain of age $a$ is in each state $s$ but has not been any state in $S_1$ in the past $T_1$ units of time:
\begin{equation}
\left[P(X(a) = s \text{ and } X(\tau)\notin S_1 \ \max\{0,a-T_1\}\leq \tau \leq a\}) \right]_{s\in S}.
\end{equation}
To derive an expression for these probabilities we consider a modified Markov chain $X_a$ which until time $\max\{0,a-T_1\}$ behaves as the original Markov chain, but at time $\max\{0,a-T_1\}$ individuals in $S_1$ are moved to the absorbing `death' state, and after $\max\{0,a-T_1\}$ all state transitions entering states in $S_1$ are now redirected to the absorbing `death' state.
For $X_a$ the transition rate matrix between states in $S$ after time $\max\{0,a-T_1\}$ is $Q$ with the rows and columns corresponding to states in $S_1$ set to zero. If we assume, without loss of generality, that the states are ordered such that states in $S_1$ are last we can write this in block matrix form as 
$\left[\begin{array}{cc} Q_1 & 0 \\ 0 & 0 \end{array}\right]$,
where is $Q_1$ is the sub-matrix of $Q$ corresponding to states in $S \setminus S_1$. Therefore 
\begin{align}
\left[P(X(a) = s \text{ and } X(\tau)\notin S_1 \ \max\{0,a-T_1\}\leq \tau \leq a) \right]_{s\in S}
&=
\left[P(X_a(a) = s)\right]_{s\in S}\\
&=
\begin{cases}
\bbP^a_1 \bs{\pi}, \ a<T_1 \\
\bbP_1^{T_1} \bbP^{a-T_1} \bs{\pi}, \ a \geq T_1,
\end{cases}
\end{align}
where $\bbP := e^Q$ and
$\bbP_1 :=
\left[\begin{array}{cc}
e^{Q_1} & 0 \\
0       & 0 \end{array}\right]
=\exp{\left[\begin{array}{cc}
Q_1 & 0 \\
0   & 0 \end{array}\right]}
\left[\begin{array}{cc}
I & 0 \\
0 & 0 \end{array}\right]$ are matrices the same size as $Q$.

Therefore the expected number of people at equilibrium which have not been in some set of states $S_1\subset S$, in the past $T_1$ units of time is the vector
\begin{equation}
\lim_{t\to \infty}\int_0^t \left[\mu(t-a)[P(X_a(a) {=} s)\right]_{s\in S} da
= \lim_{t\to \infty} \left[ \int_0^{T_1} \mu(t-a) \bbP^a_1 \bs{\pi} da + \int_{T_1}^t \mu(t-a) \bbP^T_1 \bbP^{a-T_1} \bs{\pi} da \right]
\end{equation}
If $\mu$ is constant then this simplifies to
\begin{align}
\mu \left[ \int_0^{T_1}\bbP^a_1 da + \int_{T_1}^\infty  \bbP^{T_1}_1 \bbP^{a-T_1} da \right] \bs{\pi}
&=
\mu \left[ \int_0^{T_1}\bbP^a_1 da + \int_{0}^\infty  \bbP^{T_1}_1 \bbP^{a} da \right] \bs{\pi}\\
&=\mu \left[ \bbQ_1 \left(\bbP^{T_1}_1 - I \right) -  \bbP^{T_1}_1 Q^{-1} \right] \bs{\pi}
\end{align}
where by a abuse of notation,
$\bbQ_1 := \left[\begin{array}{cc}
Q_1^{-1} & 0 \\
0   & 0 \end{array}\right]$ is a square matrix the size of $Q$.

The same reasoning can be extended to count the individuals at equilibrium which have not been in the sets of states $S_1, S_2,\dots,S_n$ in the past $T_1 > T_2 > \dots T_n$ units of time respectively. The vector of probabilities that an individual of age $a$ is in each state $s$ and satisfies these requirements is
\begin{equation}
\left[P\left(X(a) = s \text{ and } \bigcup_{i=1}^n \{ X(\tau) \notin S_i \  \ \max\{0,a-T_i\}\leq \tau \leq a\}\right) \right]_{s\in S}
= \begin{cases}
\bbP_n^a \bs{\pi}, &\ a<T_n \\
\bbP_n^{T_n} \bbP_{n-1}^{T_{n-1}-T_n} \dots \bbP_i^{T_{i} -T_{i+1}} \bbP_{i-1}^{a-T_i} \bs{\pi}, &\ T_i \leq a < T_{i-1}\\
\bbP_n^{T_n} \bbP_{n-1}^{T_{n-1}-T_n} \dots \bbP_1^{T_1 -T_2} \bbP^{a-T_1} \bs{\pi}, &\ T_1 \leq a
\end{cases}
\end{equation}
and the expected number of each type of individual in the population at equilibrium (assuming constant birth rate $\mu$) is
\begin{align*}
\mu[
  \bbQ_n (\bbP_n^{T_n} - I)
+& \bbP_n^{T_n}\bbQ_{n-1} (\bbP_{n-1}^{T_{n-1}} - I)
+ \bbP_n^{T_n}\bbP_{n-1}^{T_{n-1}-T_{n}} \bbQ_{n-2}(\bbP_{n-2}^{T_{n-2}}-I)
+ \dots \\
+& \bbP_n^{T_n}\bbP_{n-1}^{T_{n-1}-T_{n}}\dots\bbP_2^{T_2-T_3}\bbQ_1 (\bbP_1^{T_1} - I)
- \bbP_n^{T_n}\bbP_{n-1}^{T_{n-1}-T_{n}}\dots\bbP_1^{T_1-T_2}Q^{-1}
]\bs{\pi}
\end{align*}
where $\bbP_i :=
\left[\begin{array}{cc}
e^{Q_i} & 0 \\
0       & 0 \end{array}\right]$
and
$\bbQ_i := \left[\begin{array}{cc}
Q_i^{-1} & 0 \\
0   & 0 \end{array}\right]$ are matrices the same size as $Q$ with $Q_i$ being the sub-matrices of $Q$ corresponding to the states in $S \setminus \cup^n_{j=i} S_j$.

Now we have a way to calculate the incidence of (non-recurrent) CDIs which would be classified as hospital or community acquired if standard definitions were used. We will briefly illustrate this for community onset cases for the system recommended by IDSA and SHEA.

First calculate the equilibrium number of people in each class in the community that haven't been in symptomatic states ($S_1$) in the past 8 weeks ($T_1$). At this equilibrium point calculate the rate at which the transitions corresponding to onset of CDIs occur in the community, i.e. the total rate of transitions from asymptomatically colonised community states to symptomatic community states. This is the incidence of non-recurrent community-onset CDI.

The rate of non-recurrent CDI classified as community-acquired is the total rate of transitions corresponding to the onset of CDI for the equilibrium  number of people who haven't been in the hospital states ($S_1$) in the past 12 weeks ($T_1$) or any symptomatic state ($S_2$) in the past 8 weeks ($T_2$).

The rate of non-recurrent CDI classified as indeterminate is the total rate of transitions corresponding to the onset of CDI for the equilibrium  number of people who haven't been in the hospital states ($S_2$) in the past 4 weeks ($T_2$) or any symptomatic state ($S_1$) in the past 8 weeks ($T_1$) minus the rate of non-recurrent CDIs classified as community-acquired.

Finally, the rate of non-recurrent CDI classified as hospital-acquired is the rate of non-recurrent CDI minus the rate of non-recurrent CDIs classified as community-acquired or indeterminate.

The first steps of classifying of hospital-onset is similar, however the history of recurrence and previous hospitalisation is considered from the point of hospital admission. In other other words, we determine the proportion of admission which have or have not been in a hospital in the past 12 weeks etc. Given the history of patients at the point of admission, one can simply simulate the course of hospitalisations going forward and  (e.g. how many CDIs occur within the first 2 days) and classify all cases as hospital or community acquired to the history at admission and what occurs during the hospital stay.

To account for possible unreported cases of CDI in the community, once we calculate the incidence of community-onset CDIs, we then calculate (using standard Markov chain calculations for absorption probabilities) the number of community-onset cases that seek treatment in the community or in the hospital. A fraction $p_{Report}$ of patients that seek treatment in the community and all patients that do not seek treatment in the community but are admitted to hospital are considered `reported' and count towards the incidence calculation when fitting the model to Lessa et al. A small number patients seek treatment in the community and then are also admitted to hospital before recovery; for simplicity only $p_{Report}$ of these are considered reported. Patients that do not seek any form of treatment before recovering are not counted towards the `reported' incidence calculation. Note that this adjustment for `reporting' is not made when we determine whether a case is recurrent or not; that is any CDI (even if didn't seek treatment) in the patient's recent history excludes the current CDI as recurrent. This close approximation greatly simplifies the calculations for excluding recurrent cases. 

To see the true (not classified) location of acquisition for each CDI, we use a modified model which splits every \emph{C. difficile}-positive compartment into separate compartments for hospital and community-acquired. The incidence of community-acquired CDI is then the equilibrium transition rate from community-acquired, asymptomatically colonised states to the corresponding community-acquired, symptomatically colonised states.

To compare the classification system to the true incidence, we use the methods described above on the modified model. For instance, we can calculate the incidence of community-acquired, community-onset CDI that is incorrectly classified as hospital-acquired, community-onset CDI: the rate at which people in community-acquired asymptomatically colonised states (that have been in any of the hospital states in the past 4 weeks) transition into symptomatic states in the community. Note, we assume that the time and place of the onset of symptoms is always known accurately, i.e. community-onset cases are never misclassified as hospital-onset or vice versa.

\includepdf[pages={1-last}]{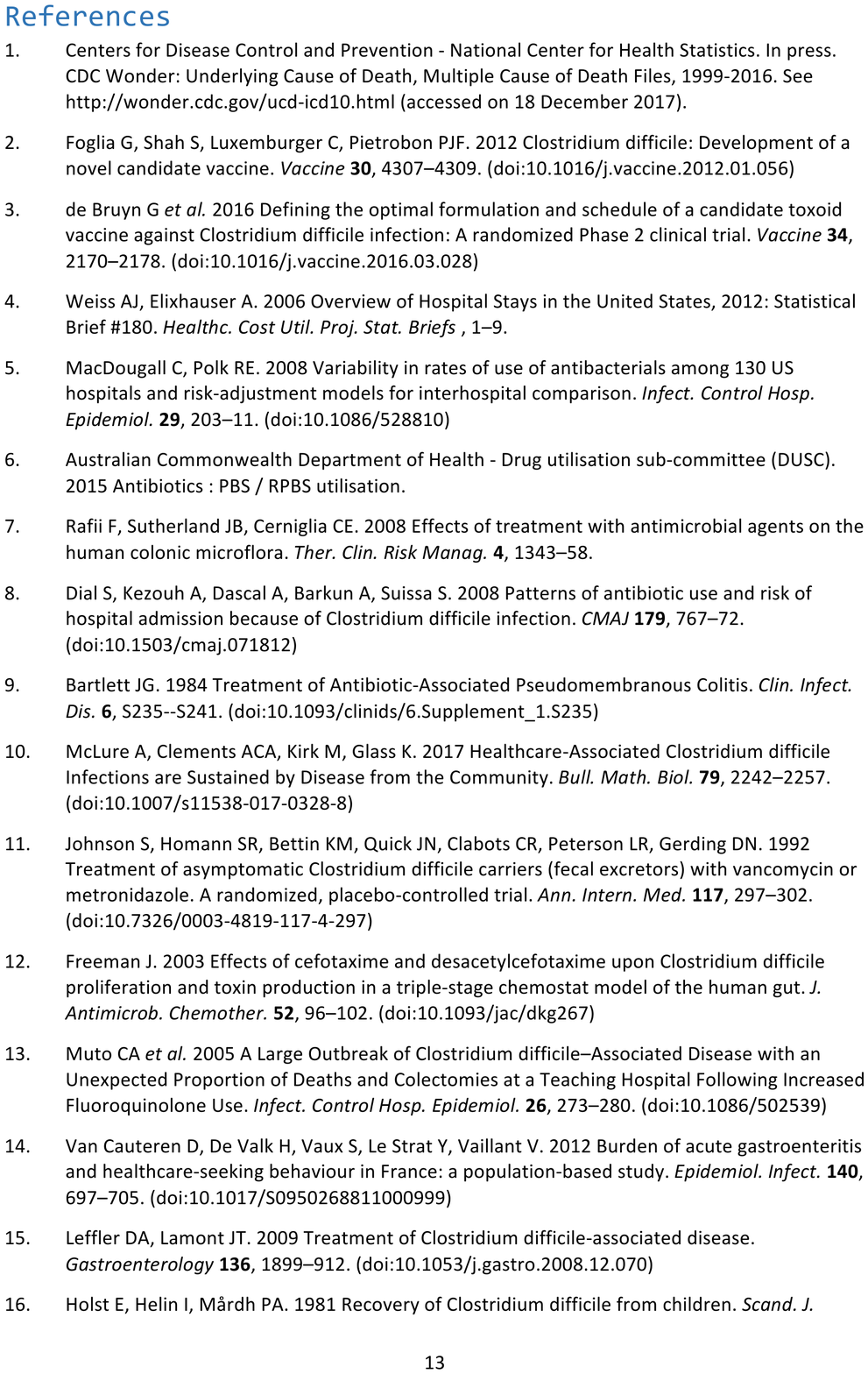}
\end{document}